\documentclass[]{spie}  

\usepackage{amsmath,amsfonts,amssymb,multirow}
\usepackage{graphicx}
\newcommand{\lt}{\left}
\newcommand{\rt}{\right}
\usepackage[colorlinks=true, allcolors=blue]{hyperref}
\usepackage{afterpage}

\title{Iterative Reconstruction of the Electron Density and Effective Atomic Number using a Non-Linear Forward Model}

\author[a]{K. Aditya Mohan}
\author[a]{Kyle M. Champley}
\author[b]{Albert W. Reed}
\author[a]{Steven M. Glenn}
\author[a]{Harry E. Martz Jr.}
\affil[a]{Lawrence Livermore National Laboratory, Livermore, CA, USA}
\affil[b]{Arizona State University, Tempe, AZ, USA}

\authorinfo{Further author information: (Send correspondence to K. A. M.)\\K. A. M.: E-mail: mohan3@llnl.gov; K. M. C.: E-mail: champley1@llnl.gov; A. W. R.: E-mail: awreed@asu.edu; S. M. G.: E-mail: glenn21@llnl.gov; H. E. M.: E-mail: martz2@llnl.gov.\\
}

\begin{document} 
\maketitle

\begin{abstract}
For material identification, characterization, and quantification, it is useful to estimate system-independent material properties that do not depend on the detailed specifications of the X-ray computed tomography (CT) system such as spectral response. System independent $\rho_e$ and $Z_e$ (SIRZ) refers to a suite of methods for estimating the system independent material properties of electron density ($\rho_e$) and effective atomic number ($Z_e$) of an object scanned using dual-energy X-ray CT (DECT). The current state-of-the-art approach, SIRZ-2, makes certain approximations that lead to inaccurate estimates for large atomic numbered ($Z_e$) materials. In this paper, we present an extension, SIRZ-3, which iteratively reconstructs the unknown $\rho_e$ and $Z_e$ while avoiding the limiting approximations made by SIRZ-2. Unlike SIRZ-2, this allows SIRZ-3 to accurately reconstruct $\rho_e$ and $Z_e$ even at large $Z_e$.  SIRZ-3 relies on the use of a new non-linear differentiable forward measurement model that expresses the DECT measurement data as a direct analytical function of $\rho_e$ and $Z_e$. Leveraging this new forward model, we use an iterative optimization algorithm to reconstruct (or solve for) $\rho_e$ and $Z_e$ directly from the DECT data. 
Compared to SIRZ-2, we show that the magnitude of performance improvement using SIRZ-3 increases with increasing 
values for $Z_e$.
\end{abstract}

.

\section{INTRODUCTION}
\label{sec:intro}  

X-ray computed tomography (CT) is widely used to reconstruct
the spatially varying linear attenuation coefficient (LAC) of an object \cite{Martz_XCTBook_2016}.
The spectral response function that is dependent on the X-ray source and the detector is used to 
quantify the variation in X-ray intensity
over the broad range of emitted X-ray energies.
Most CT systems have differing spectral response functions that causes the reconstructed LAC to be dependent 
on the spectral response of the X-ray scanner. 
Hence, it is not possible to compare the 
LAC values reconstructed using different X-ray scanners. 
Also, the estimated LAC does not provide direct information on the density and
atomic composition of an object \cite{Champley_SIRZ2_2019}.

Dual-energy X-ray computed tomography (DECT) is useful to estimate 
system independent properties that are independent of the CT system spectral response and other system specifications.
Our laboratory based micron scale DECT system scans the same object twice using two different spectra.
Using DECT scans, an approach called SIRZ was proposed \cite{Azevedo_SIRZ_2016}
to estimate the system independent properties of electron density, $\rho_e$, 
and effective atomic number, $Z_e$.
The electron density, $\rho_e$, is a direct measure of the object's density
and the effective atomic number, $Z_e$, is useful to predict the atomic composition.
In Champley at al.\cite{Champley_SIRZ2_2019}, the authors presented the SIRZ-2 approach
that improved upon the original SIRZ to produce more accurate estimates for $\rho_e$ and $Z_e$.
SIRZ-2 can be broadly split into four different steps: 
\begin{enumerate}
\item Preprocessing steps that include detector gain correction, detector deblur, and scatter correction.
\item Dual-energy basis decomposition to estimate projections of the object at two different synthesized monochromatic basis (SMB) energies.
\item Filtered back-projection to reconstruct the LAC values at the two SMB energies,
\item Estimation of electron density and effective atomic number from the LACs at the two SMB energies.
\end{enumerate}
The second and fourth steps of SIRZ-2 are based on simplifying approximations that limit its accuracy.
 
In this paper, we present a new approach called SIRZ-3
to reconstruct the system independent properties of $\rho_e$ and $Z_e$
while avoiding the approximations made in SIRZ-2.
SIRZ-3 replaces the steps 2, 3, and 4 of SIRZ-2 using a single
iterative optimization algorithm.
Features of the SIRZ-3 method include:
\begin{itemize} 
\item Non-linear differentiable forward model that expresses the measured DECT data
as a direct function of the object's spatial distribution of $\rho_e$ and $Z_e$.
\item Formulation of an error function that quantifies the deviation of measured data from 
the output of the forward model given the unknowns, $(\rho_e, Z_e)$. 
\item Method for reconstructing $(\rho_e, Z_e)$ from the measured data by minimizing the error function using the limited memory Broyden–Fletcher–Goldfarb–Shanno (L-BFGS) solver \cite{Liu_LBFGS_1989, Nocedal_LBFGS_1980}.
\end{itemize}

\section{Background}
\label{sec:back}
The LAC
quantifies the magnitude of X-ray attenuation
within an object.
The LAC varies as a function of the position inside the object, density, atomic composition, 
and X-ray energy.
The dependence of LAC, denoted as $\mu\lt(Z, \rho_e, E\rt)$, on the electron density, $\rho_e$, 
and atomic number, $Z$, is given by,
\begin{equation}
\label{eq:lacvsZE}
\mu\lt(Z, \rho_e, E\rt) = \rho_e \sigma_e (Z, E),
\end{equation}
where $E$ is the X-ray energy, $\sigma_e(\cdot)$ is the total X-ray electronic cross-section.\cite{Champley_SIRZ2_2019} 
Equation \eqref{eq:lacvsZE} is only defined for integer valued $Z$
that correspond to pure elements. 

In order to represent compounds or mixtures of elements, 
the cross-section for non-integer $Z_e$ is defined as follows \cite{Azevedo_SIRZ_2016}, 
\begin{equation}
\label{eq:xsecvsZeE}
\sigma_e\lt(Z_e, E\rt) = \lt(1-\delta\rt) \sigma_e\lt(Z', E\rt) + \delta\sigma_e\lt(Z'+1, E\rt),
\end{equation}
where $Z'=\lfloor Z_e\rfloor$, and $\delta = \lt(Z_e-Z'\rt)$.
Note that $Z_e$ for a given compound or mixture is not a fundamental physical constant 
and there are several competing techniques to define $Z_e$.
We adopt the definition of $Z_e$ from Champley et al. \cite{Champley_SIRZ2_2019}
since it is a function of the X-ray cross-section with superior system invariance properties.
The $Z_e$ for a compound or mixture is defined as,
\begin{equation}
\label{eq:effZ}
Z_e = \arg\min_{z} \lt\lbrace \int_E S(E) \lt[\exp\lt(-M\sigma_e(z, E)\rt)-\exp\lt(-M\sigma_c(E)\rt)\rt]^2 dE\rt\rbrace,
\end{equation}
where $\sigma_c(E)$ is the tabulated\cite{OSTI_97} electron cross-section of the material,
$S(E)$ is the spectral response at energy $E$, and 
$M$ is the areal electron density.
This definition in \eqref{eq:effZ} is demonstrated to be 
relatively insensitive to $S(E)$ and the X-ray energy range of interest (30-200keV).
For additional details on the definition of $Z_e$
and its estimation, the reader may refer to Champley et al. \cite{champley_livermore_2022}.

\section{FORWARD MODEL}
In this section, we formulate a forward model that mathematically
expresses the measured DECT data as a function of the object properties, $\rho_e$ and $Z_e$.
This forward model is necessary to solve the inverse problem
of reconstructing the object's $\rho_e$ and $Z_e$.

The forward model is derived in discrete coordinate space.
Let $\rho_{e,j}$ and $Z_{e,j}$ represent the $\rho_e$ and $Z_e$ values respectively 
at the $j^{th}$ voxel inside the object.
Then, the LAC at voxel $j$ is given by $\mu_{j,k} = \rho_{e,j} \sigma_k\lt(Z_{e,j}\rt)$,
where $\sigma_k\lt(Z_{e,j}\rt) = \sigma_e\lt(Z_{e,j}, E_k\rt)$ from \eqref{eq:xsecvsZeE}
and $E_k$ is the X-ray energy at the $k^{th}$ energy bin.
The linear projections (line-integration) of the LAC is expressed as,
\begin{equation}
p_{i,k} = \sum_j A_{i,j} \mu_{j,k},
\end{equation}
where $A_{i,j}$ is the element along the $i^{th}$ row and $j^{th}$ column of the matrix $A$. 
At the detector, the measurement is the average transmission function $\exp\lt(-p_{i,k}\rt)$
weighted by the spectral density function $S_{k}$.
Hence, the effective transmission is given by,
\begin{equation}
\label{sec:wttrans}
t_{i} = \sum_k S_{k} \exp\lt(-p_{i,k}\rt),
\end{equation}
where $S_{k}$ is the spectral density at the $k^{th}$ energy bin ($S_{k}$ is a discretized version of $S(E)$ used in \eqref{eq:effZ})
such that $\sum_k S_k = 1$.
The spectral density function $S_k$ quantifies both
the spectral density of the X-ray source and the spectral response of the detection system. 
We compute the negative logarithm of \eqref{sec:wttrans},
which gives us the measurement data in attenuation space, $-\log\lt(t_i\rt)$.

For DECT, the complete forward model is, 
\begin{align}
\tilde{y}^L_{i} & = -\log\lt(\sum_k S^L_{k} \exp\lt\lbrace-\sum_j A_{i,j} \rho_{e,j} \sigma_k\left(Z_{e,j}\right) \rt\rbrace\rt), \label{eq:forwmodL}\\
\tilde{y}^H_{i} & = -\log\lt(\sum_k S^H_{k} \exp\lt\lbrace-\sum_j A_{i,j} \rho_{e,j} \sigma_k\left(Z_{e,j}\right) \rt\rbrace\rt), \label{eq:forwmodH}
\end{align}
where $S_k^{L}$ is the spectral density for
X-rays at the low-energy spectrum,
$S_k^H$ is the spectral density at the high-energy spectrum, and $\tilde{y}^L_i$ and $\tilde{y}^H_{i}$ are
the forward model predictions for the 
low-energy and high-energy measurements
respectively.
To ensure differentiability of the forward model
in \eqref{eq:forwmodL} and \eqref{eq:forwmodH},
we need analytical derivatives for $\sigma_k\lt(Z_{e,j}\rt)$ with respect to $Z_{e,j}$.
This derivative is given by, 
\begin{equation}
\frac{\partial \sigma_k\lt(Z_{e,j}\rt)}{\partial Z_{e,j}} =
\sigma_k\lt(Z'_{e,j}+1\rt) - \sigma_k\lt(Z'_{e,j}\rt),
\end{equation}
where $Z'_{e,j}=\lfloor Z_{e,j}\rfloor$ and $\lt\lfloor\cdot\rt\rfloor$ is the floor function.

\section{SIRZ-3: RECONSTRUCTION BY OPTIMIZATION}
\label{sec:recalg}
SIRZ-3 is a framework to solve the problem of reconstructing $\rho_{e,j}$ and $Z_{e,j}$
from measurements $y^L_i$ and $y^H_i$ using a forward model that accurately defines an analytical relation between $\lt(y^L_i, y^H_i\rt)$ and $\lt(\rho_{e,j}, Z_{e,j}\rt)$ for all $i$ and $j$.
SIRZ-3 adopts an iterative approach, where
the solution for $\lt(\rho_{e,j}, Z_{e,j}\rt)$ is iteratively refined such that the output
of the forward model (whose input is the current estimate for $\rho_{e,j}$ and $Z_{e,j}$) gets progressively closer to the measurements $\lt(y^L_i, y^H_i\rt)$.
In addition to a forward model, the solution
for $\lt(\rho_{e,j}, Z_{e,j}\rt)$ may also be required
to satisfy certain sparsity enforcing regularization 
criteria using a prior model.
In this paper, however, we do not use a prior model. 


The analytical relation for the forward model used in this paper is shown in equations \eqref{eq:forwmodL} and \eqref{eq:forwmodH}.
The accuracy of SIRZ-3 is dependent on the accuracy of this analytical relation 
that mathematically describes the measurement process of our X-ray system.
Future efforts to improve SIRZ-3 may potentially focus on more accurate forward models
that improve upon equations \eqref{eq:forwmodL} and \eqref{eq:forwmodH} or more accurate
preprocessing algorithms so that 
\eqref{eq:forwmodL} and \eqref{eq:forwmodH} does accurately model the data.

To formulate the reconstruction algorithm, we define the following error function,
\begin{equation}
\label{eq:errfunc}
E\lt(y^L, y^H; \rho_e,Z_e\rt) = \sum_i w^L_i\lt(y^{L}_i-\tilde{y}^L_i\rt)^2 +  \sum_i w^H_i\lt(y^{H}_i-\tilde{y}^H_i\rt)^2,
\end{equation}
where $y^L_i$ and $y^H_i$ are the attenuation space measurements at the low-energy
and high-energy spectra respectively, and $\tilde{y}_i^L$ and $\tilde{y}_i^H$
are the forward model predictions from \eqref{eq:forwmodL} and \eqref{eq:forwmodH} respectively.
The weights for the penalty terms are given by $w^L_i=\exp\lt(-y^L_i\rt)/N$ and $w^H_i=\exp\lt(-y^H_i\rt)/N$, where $N$ is the number
of sinogram pixels summed over all the views.
Substituting \eqref{eq:forwmodL} and \eqref{eq:forwmodH} in \eqref{eq:errfunc}, we get,
\begin{multline}
\label{eq:suberrfunc}
E\lt(y^L, y^H; \rho_e,Z_e\rt) = 
\sum_i w^L_i\lt(y^L_{i}+\log\lt[\sum_k S^L_{k}
\exp\lt\lbrace-\sum_j A_{i,j} \rho_{e,j} \sigma_k\lt(Z_{e,j}\rt)\rt\rbrace \rt]\rt)^2 \\
+ \sum_i w^H_i\lt(y^H_{i}+\log\lt[\sum_k S^H_{k}
\exp\lt\lbrace-\sum_j A_{i,j} \rho_{e,j} \sigma_k\lt(Z_{e,j}\rt)\rt\rbrace \rt]\rt)^2,
\end{multline}
where $Z_e$ is a vector of all $Z_{e,j}$ voxel values, $\rho_e$ is a vector of all $\rho_{e,j}$ voxel values.
Note that the projection matrix entries $A_{i,j}$
are stored in a  memory-efficient sparse representation
that only stores the non-zero values and its locations. 
We used LTT \cite{champley_livermore_2022}
to generate $A_{i,j}$.
Then, the reconstruction of SIRZ-3 is obtained by solving,
\begin{equation}
\label{eq:optfunc}
\lt(\hat{\rho}_e,\hat{Z}_e\rt) = \arg\min_{\rho_e,Z_e} E\lt(y^L, y^H; \rho_e,Z_e\rt), \text{ s.t. } \rho^{min}_{e} \leq \rho_{e,j} \leq \rho_{e}^{max} \text{ and } Z_{e}^{min}\leq Z_{e,j} \leq Z_{e}^{max},
\end{equation}
where $\hat{Z}_e$ and $\hat{\rho}_e$ are reconstructions of $Z_e$ and $\rho_e$ respectively,
$\rho_{e}^{min}$ and  $\rho_{e}^{max}$ are the lower
and upper limits for $\rho_{e,j}$ respectively, 
and $Z_{e}^{min}$ and  $Z_{e}^{max}$ are the lower
and upper limits for $Z_{e,j}$ respectively.
In this paper, we set
$Z_{e}^{min} = 1$, $Z_{e}^{max}=118$, $\rho_{e}^{min}=0$, and
$\rho_{e}^{max}=9.018507\times 10^{-3}$electrons$\times$mol/mm$^3$.

In order to improve conditioning and ensure
fast convergence, 
we optimize z-score normalized versions of 
$Z_e$ and $\rho_e$.
Let $Z_{e}^{avg}$ and $Z_{e}^{std}$ be the mean
and standard deviation of the SIRZ-2 reconstructed voxel values for $Z_e$. 
Similarly, let $\rho_{e}^{avg}$ and $\rho_{e}^{std}$ be the mean
and standard deviation of the SIRZ-2 reconstructed voxel values for $\rho_e$. 
Then, we solve,
\begin{multline}
\label{eq:optfuncnm}
\lt(\hat{\rho}^{nm}_e,\hat{Z}^{nm}_e\rt) = \arg\min_{\rho^{nm}_e,Z^{nm}_e} E\lt(y^L, y^H; \rho_{e}^{std}\rho^{nm}_e+\rho_{e}^{avg}, Z_{e}^{std}Z^{nm}_e+Z_{e}^{avg}\rt), \\
\text{ s.t. } \rho_{e}^{min} \leq \rho_{e}^{std}\rho^{nm}_{e,j}+\rho_{e}^{avg} \leq \rho_{e}^{max} \text{ and } Z_{e}^{min}\leq Z_{e}^{std}Z^{nm}_{e,j}+Z_{e}^{avg} \leq Z_{e}^{max},
\end{multline}
where we use vector-scalar multiplication\footnote{Vector-scalar multiplication is point wise multiplication of a scalar to every element of a vector.} and vector-scalar addition\footnote{Vector-scalar addition is point wise addition of a scalar to every element of a vector.} within the error function $E\lt(\cdot\rt)$.
The SIRZ-3 reconstruction is given by,
\begin{equation}
\label{eq:recZerhoe}
\hat{Z}_e = Z_{e}^{std}\hat{Z}^{nm}_e+Z_{e}^{avg} \text{ and } \hat{\rho}_e = \rho_{e}^{std}\,\hat{\rho}^{nm}_e+\rho_{e}^{avg}.
\end{equation}

We use the python programming language 
and pytorch framework for implementing and solving
\eqref{eq:optfuncnm}.
Pytorch's algorithmic differentiation is used to 
compute the partial derivatives of $E\lt(\cdot\rt)$ with respect to $Z^{nm}_{e,j}$ and $\rho^{nm}_{e,j}$.
We solve equation \eqref{eq:optfuncnm} using the L-BFGS 
\cite{Nocedal_LBFGS_1980, Liu_LBFGS_1989} optimization algorithm. 
Upon convergence and using \eqref{eq:recZerhoe}, L-BFGS yields the reconstruction of the electron
density, $\hat{\rho}_e$, and the effective atomic number, $\hat{Z}_e$.
We use the LBFGS implementation by Shi et al. \cite{lbfgs_hjmshi}.
For LBFGS, we use a history size of $64$, 
weak Wolfe line search, and a convergence criteria 
that stops the LBFGS iterations when the percentage change in both $Z_e$ and $\rho_e$ goes below $0.2\%$ for $10$ consecutive iterations.
Note that it is important to 
have a sufficiently strong 
convergence criteria along
with sufficiently large history size to ensure convergence.
We use the 
SIRZ-2 estimates for $Z_e$ and $\rho_e$  to initialize the
optimization in \eqref{eq:optfuncnm}.

\makeatletter
\define@key{Gin}{simsz}[true]{%
    \edef\@tempa{{Gin}{height=1.15in, keepaspectratio=true}}%
    \expandafter\setkeys\@tempa
}
\makeatother
\begin{figure*}[!h]
\begin{center}
\begin{tabular}{cccc}
\includegraphics[simsz]{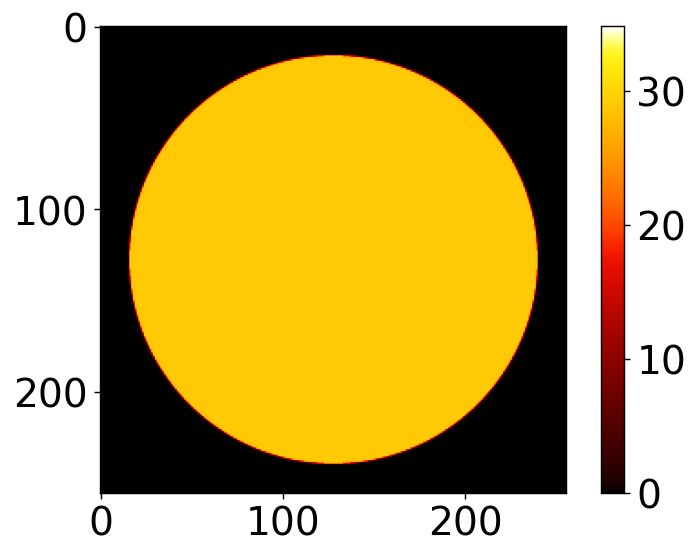} & 
\includegraphics[simsz]{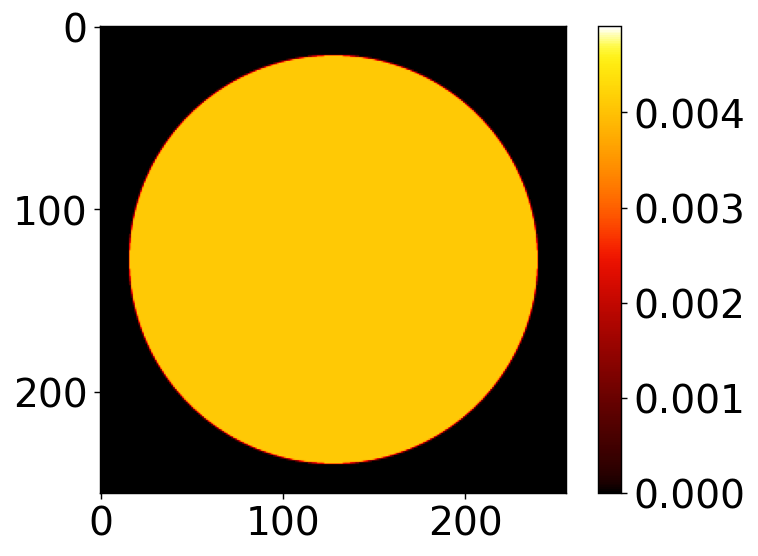} &
\includegraphics[simsz]{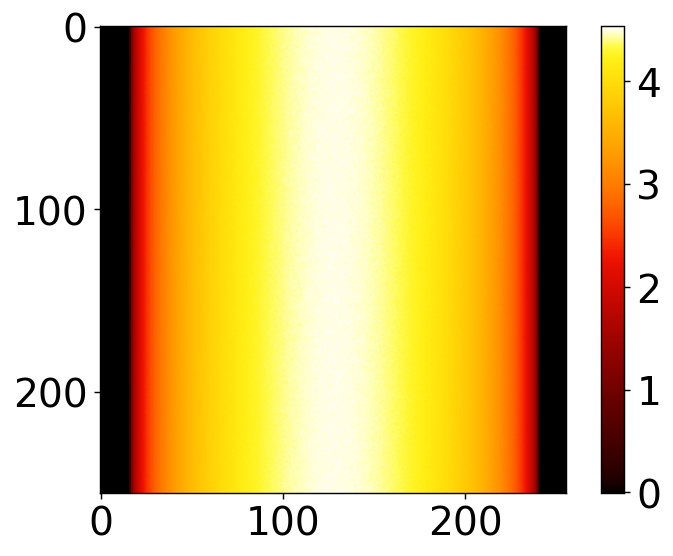} &
\includegraphics[simsz]{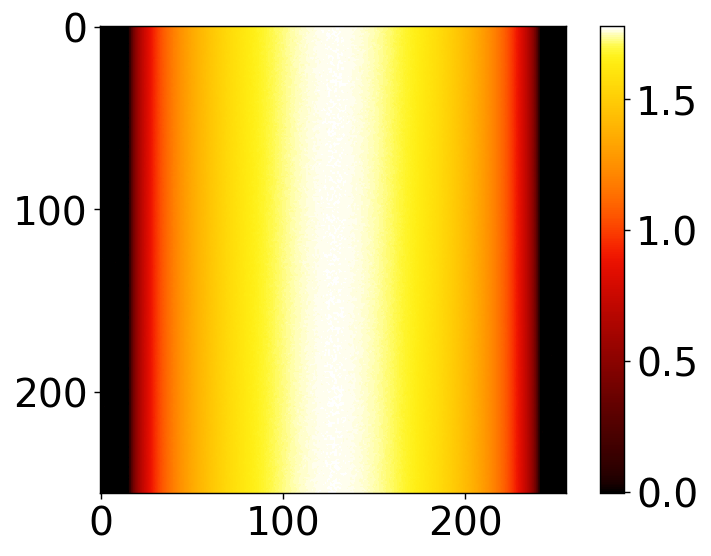} \\
(a) Ground-Truth $Z_e$ & (b) Ground-Truth $\rho_e$  & (c) Low-Energy Sinogram & (d) High-Energy Sinogram \\
\includegraphics[simsz]{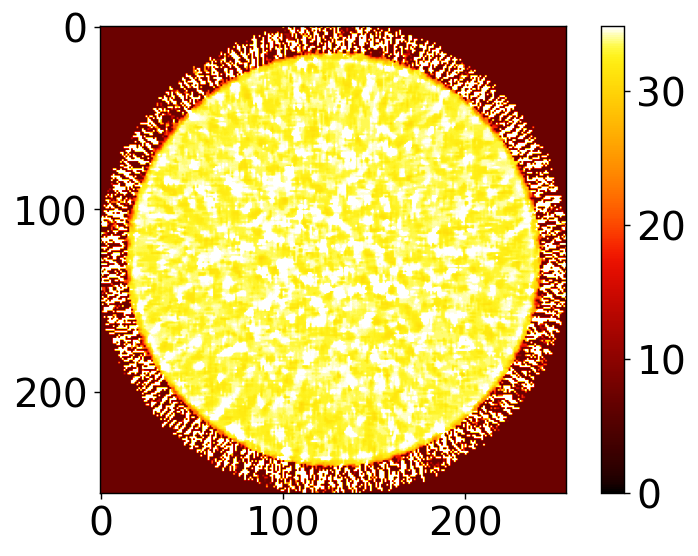} &
\includegraphics[simsz]{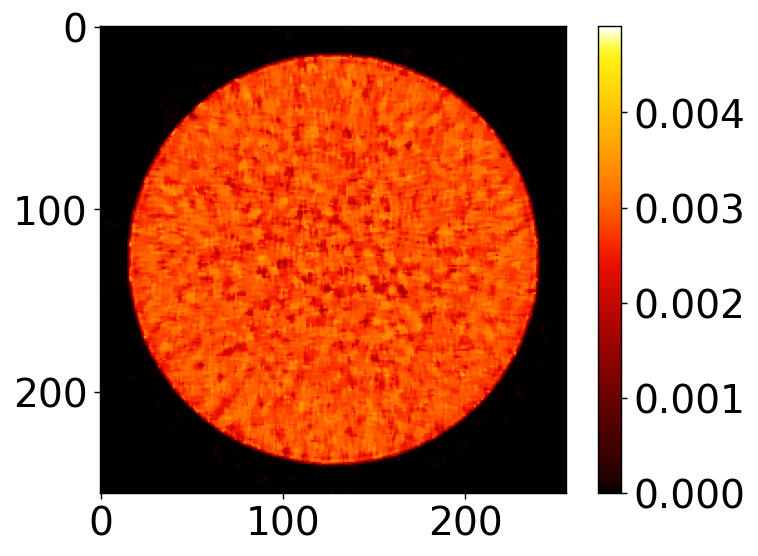} &
\includegraphics[simsz]{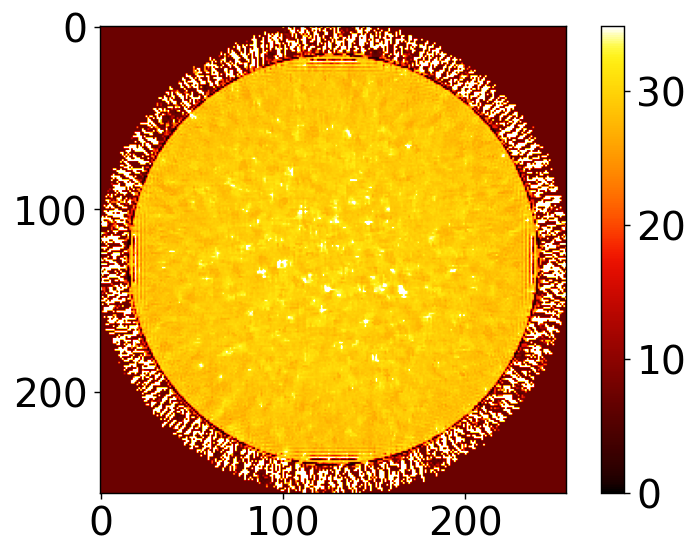} &
\includegraphics[simsz]{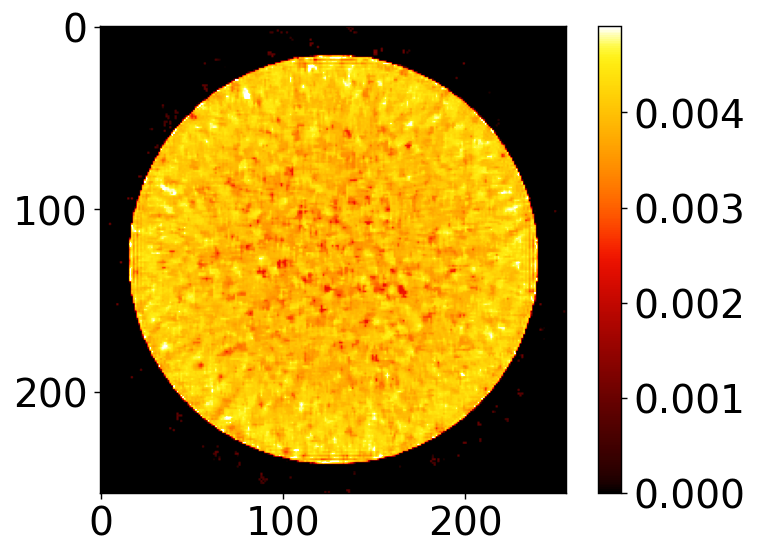} \\
(e) SIRZ-2 $Z_e$ & (f) SIRZ-2 $\rho_e$ & 
(g) SIRZ-3 $Z_e$ & (h) SIRZ-3 $\rho_e$ \\
Relative Error: $15.8\%$ & Relative Error: $-30.2\%$ & Relative Error: $1.0\%$ & Relative Error: $-1.91\%$\\
Relative RMSE: $16.36\%$ & Relative RMSE: $30.81\%$ & Relative RMSE: $4.94\%$ & Relative RMSE: $8.82\%$\\
\end{tabular}
\end{center}
\caption{\label{fig:simres}
($Z_e, \rho_e$) reconstruction of a $4$mm diameter copper (Cu) disc with a maximum attenuation of $4.5$ at the low-energy X-ray spectrum. (a, b) show the ground-truth images. (c, d) show the sinograms at the low-energy and high-energy spectra. The $y-$axis in 
(c,d) is along the view dimension. 
(e, f) and (g, h) show the SIRZ-2 and SIRZ-3 reconstructions respectively. Compared to SIRZ-2, SIRZ-3 reduces both the relative error (equation \eqref{eq:relerr}) and relative RMSE (equation \eqref{eq:relrmse}) in $Z_e$ and $\rho_e$ reconstructions.
The high relative error associated with SIRZ-2 is also observed by visually 
comparing (e, f) with (a,b).
The noisy ring-like artifacts around the $Z_e$ reconstruction of the copper disk is due to the near
zero density of the surrounding air. 
The unit for $\rho_e$ is electrons$\times$mol/mm$^3$.
}
\end{figure*}

\begin{figure*}[!h]
\begin{center}
Comparison of relative errors (\%) in $Z_e$ as defined in equation \eqref{eq:relerr}\\
\begin{tabular}{ccc}
\hspace{-0.1in}
\includegraphics[height=3.9in]{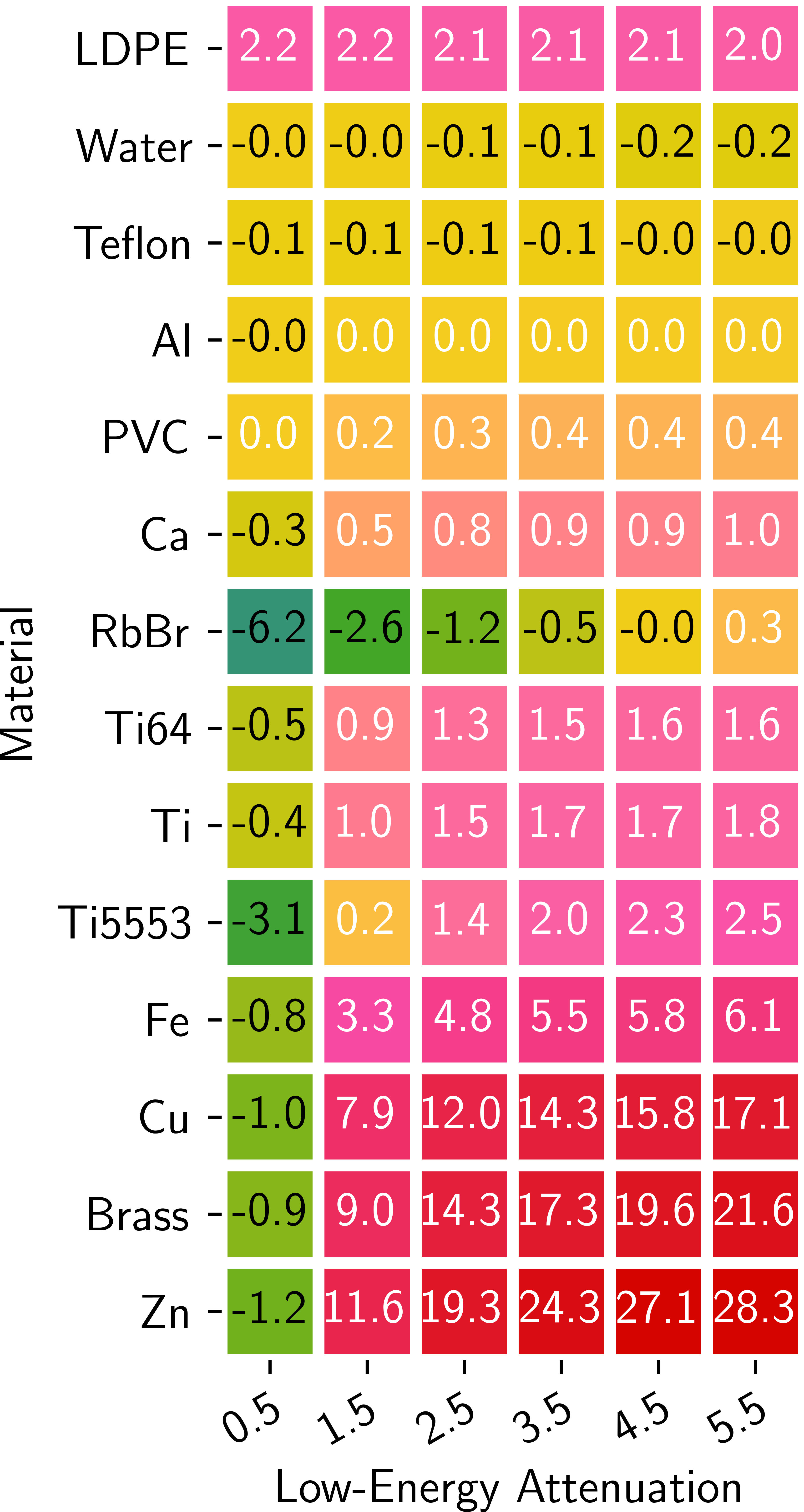} & 
\includegraphics[height=3.9in]{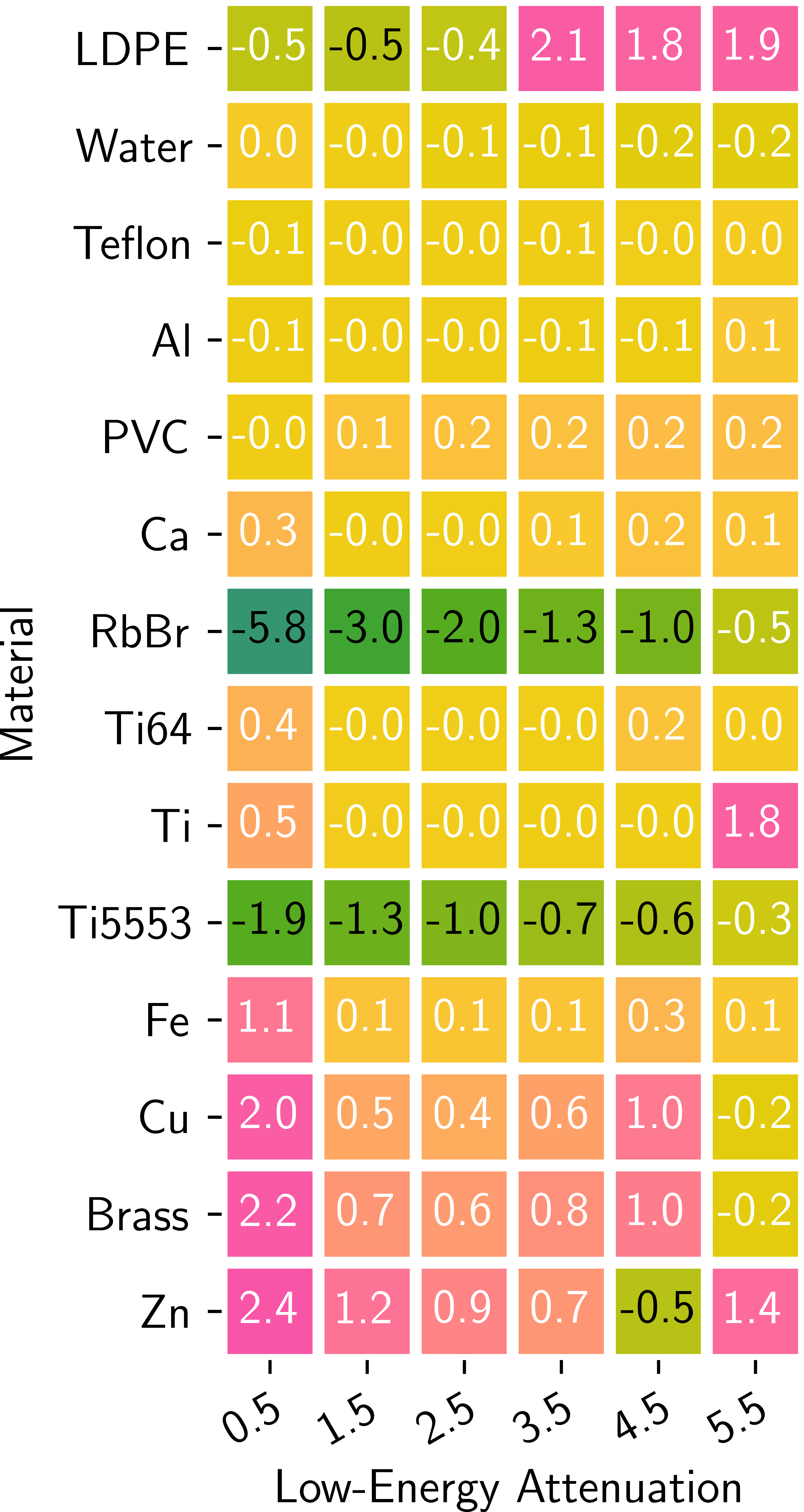} &
\includegraphics[height=3.9in]{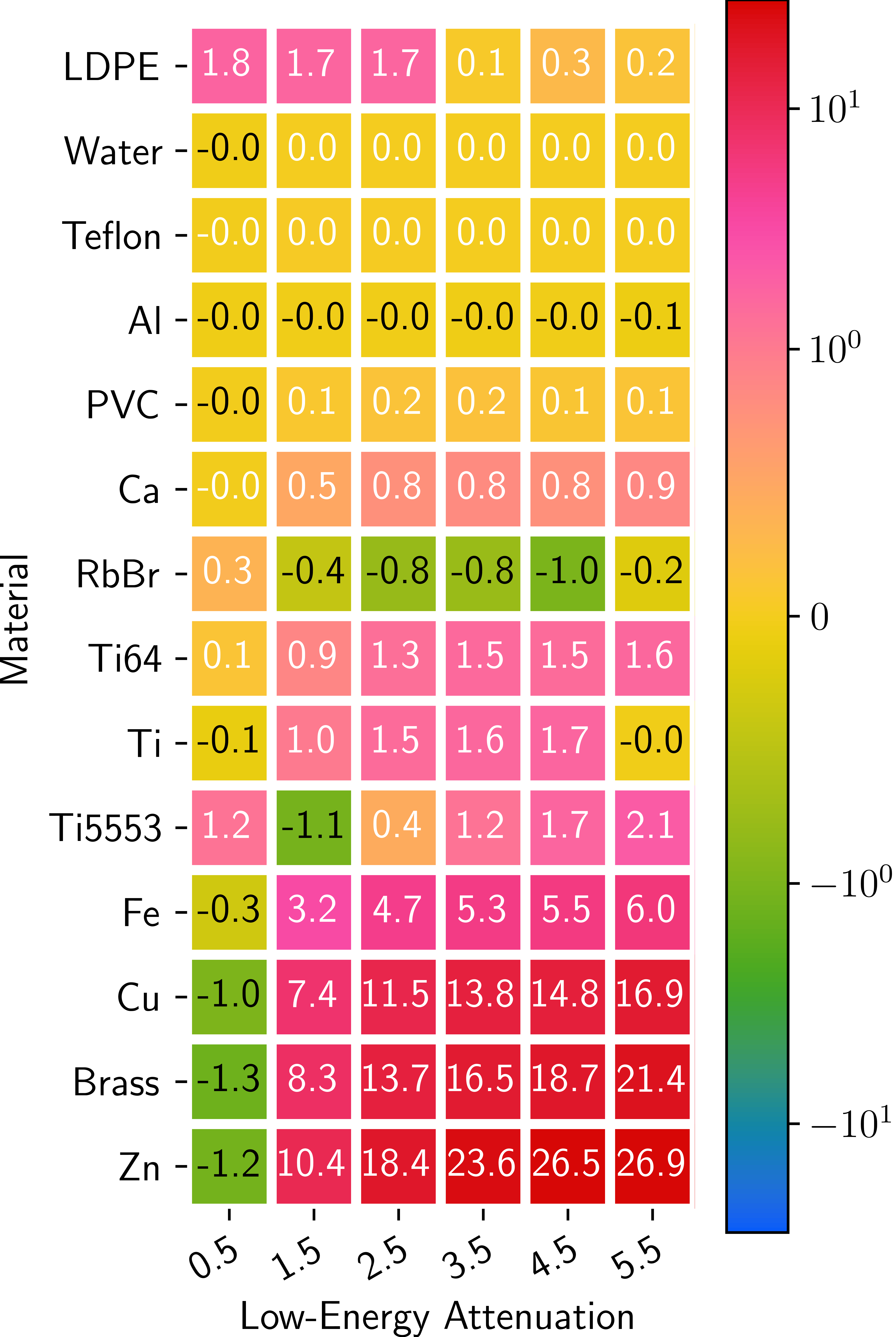}\vspace{0.05in}\\
(a) SIRZ-2 Relative Error, & (b) SIRZ-3 Relative Error,  & (c) Difference Relative Error, \\
$RE_{SIRZ-2}^{Z_e}$ & $RE_{SIRZ-3}^{Z_e}$
& $|RE_{SIRZ-2}^{Z_e}| - |RE_{SIRZ-3}^{Z_e}|$ \\
\end{tabular}
\end{center}
\caption{\label{fig:Zeerrvsmatatten} Heat map of the relative error (equation \eqref{eq:relerr}) in reconstruction of $Z_e$ for various materials and maximum low-energy attenuation values, $y^{max}_L$. (a) and (b) show the relative errors for SIRZ-2 and SIRZ-3 respectively. (c) shows the difference between the absolute values of the SIRZ-2 relative error in (a) and SIRZ-3 relative error in (b). The rows in the heat map are sorted according to the $Z_e$ value. In general, compared to SIRZ-2, the reduction in relative error of $Z_e$ using SIRZ-3 improves with increasing values for $Z_e$.
When the absolute value of the $Z_e$ relative errors with either SIRZ-2 or SIRZ-3 is greater than $1\%$, SIRZ-3 improves upon SIRZ-2 in $80\%$ of the simulated cases in the heat map. } 
\end{figure*}

\begin{figure*}[!h]
\begin{center}
Comparison of relative errors (\%) in $\rho_e$ as defined in equation \eqref{eq:relerr}\\
\begin{tabular}{ccc}
\hspace{-0.1in}\includegraphics[height=3.9in]{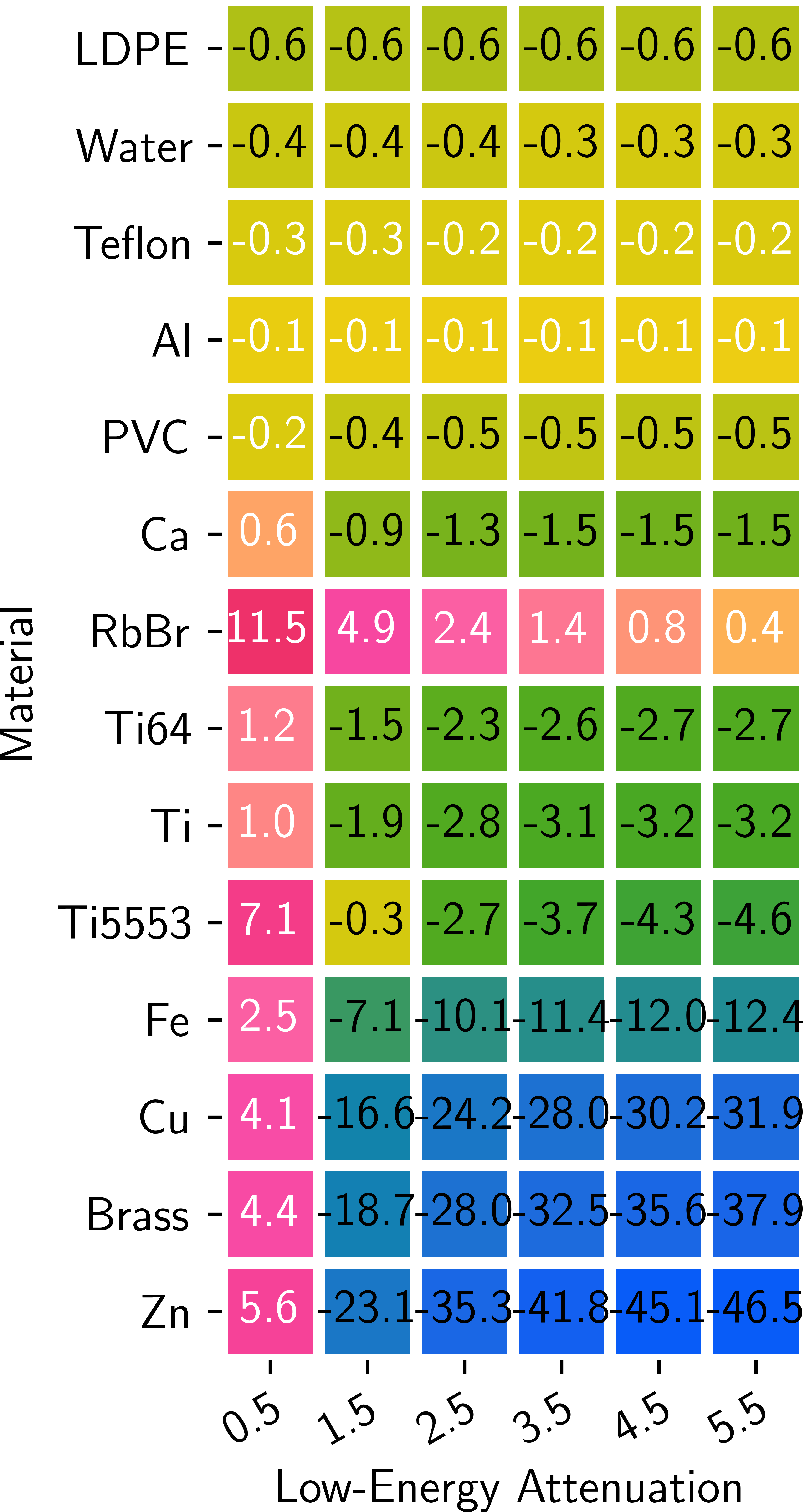} & 
\includegraphics[height=3.9in]{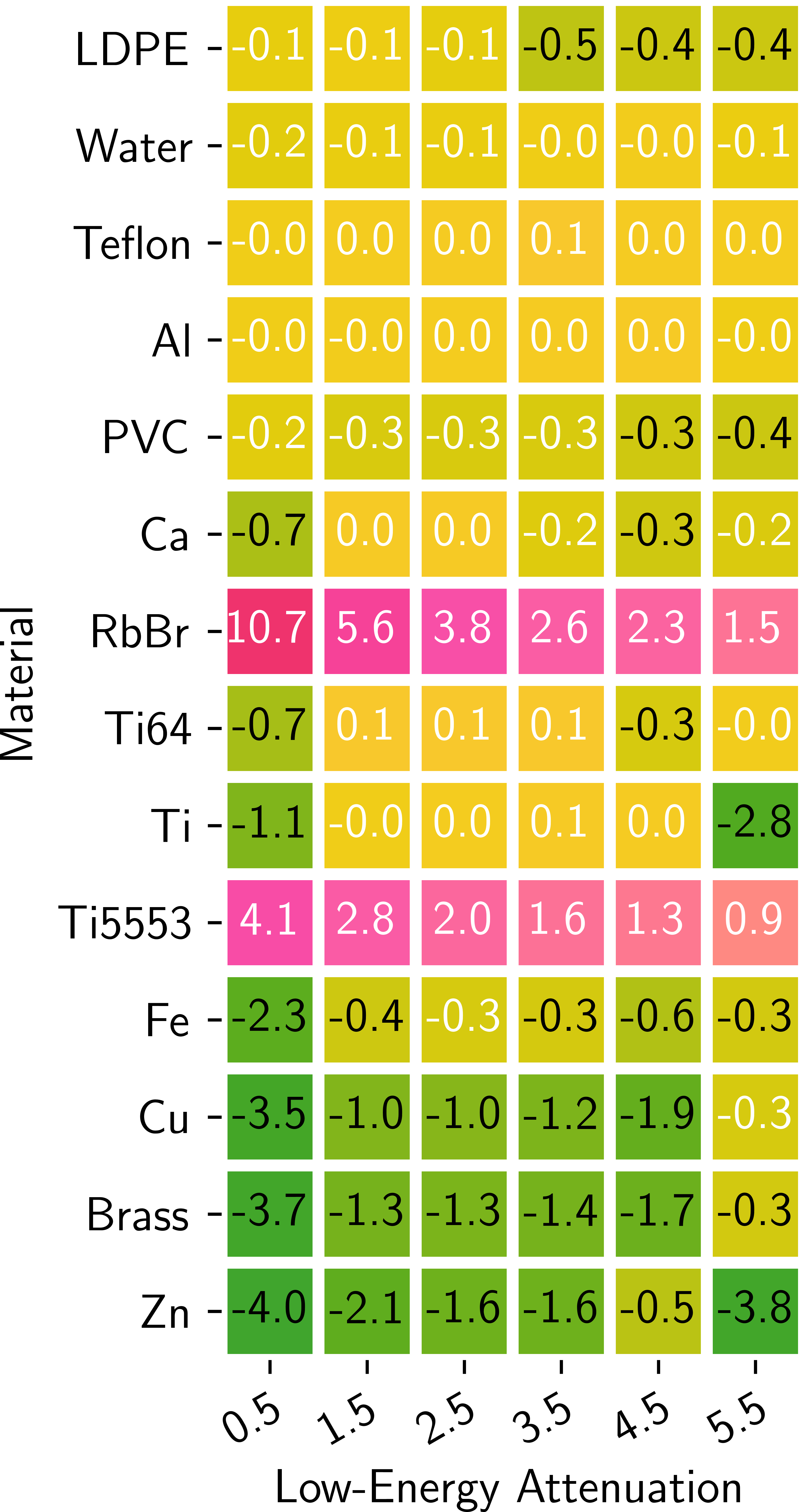} &
\includegraphics[height=3.9in]{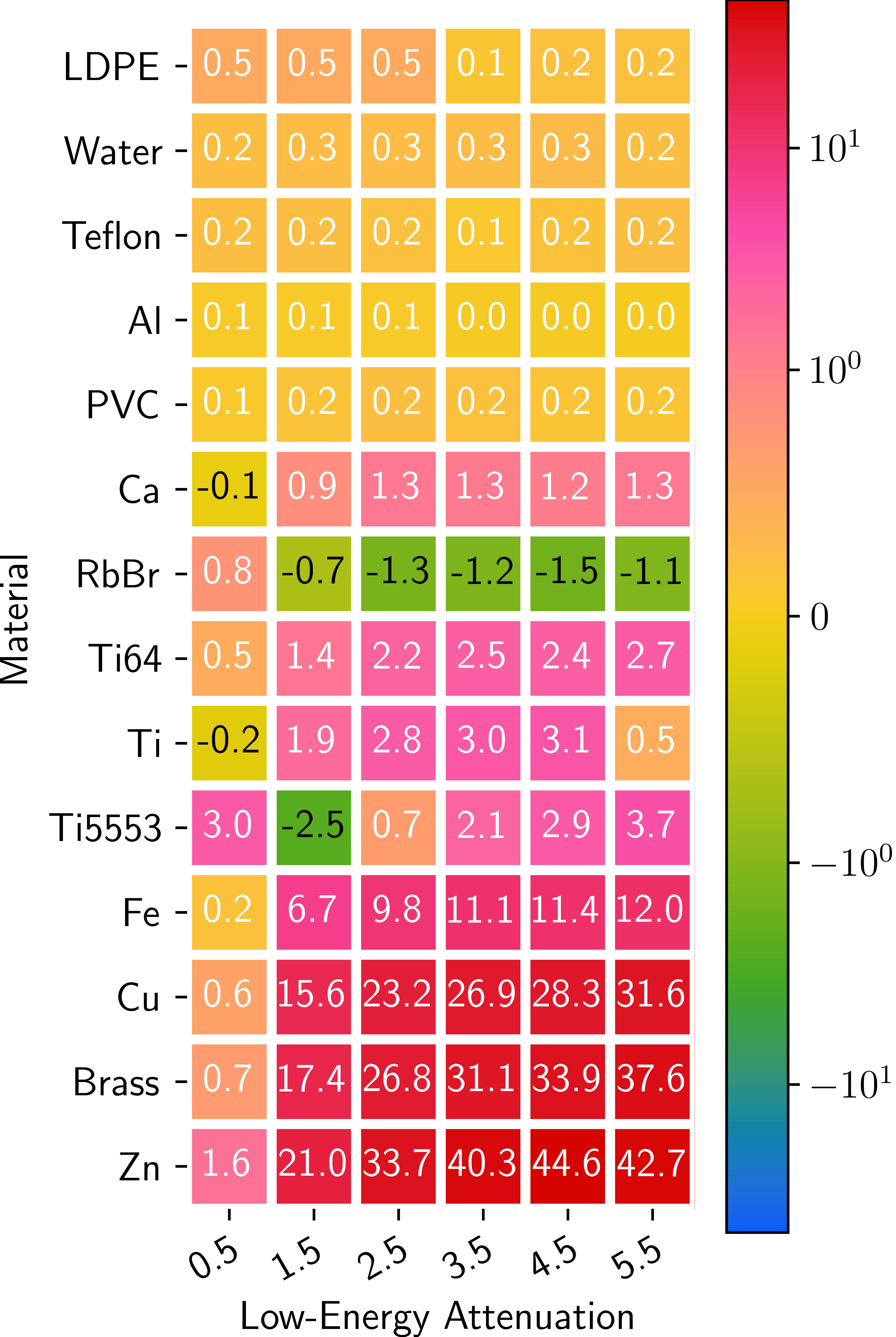}\\
(a) SIRZ-2 Relative Error, & (b) SIRZ-3 Relative Error,  & (c) Difference Relative Error, \\
$RE_{SIRZ-2}^{\rho_e}$ & $RE_{SIRZ-3}^{\rho_e}$
& $|RE_{SIRZ-2}^{\rho_e}| - |RE_{SIRZ-3}^{\rho_e}|$ \\
\end{tabular}
\end{center}
\caption{\label{fig:rhoeerrvsmatatten} Heat map of the relative error (equation \eqref{eq:relerr}) in reconstruction of $\rho_e$ for various materials and maximum low-energy attenuation values. (a) and (b) show the relative errors (equation \eqref{eq:relerr}) for SIRZ-2 and SIRZ-3 respectively. (c) shows the difference between the absolute values of the SIRZ-2 relative error in (a) and SIRZ-3 relative error in (b). The rows in the heat map are sorted according to the $Z_e$ value. In general, compared to SIRZ-2, the reduction in relative error of $\rho_e$ using SIRZ-3 improves with increasing values for $Z_e$.
When the absolute value of the $\rho_e$ relative errors with either SIRZ-2 or SIRZ-3 is greater than $1\%$, SIRZ-3 improves upon SIRZ-2 in $87\%$ of the simulated cases in the heat map. }
\end{figure*}

\section{Results}
Using simulated data, we compared the performance
of SIRZ-2 and SIRZ-3 for a wide range of materials and object thicknesses.
We chose various materials including
pure elements, compounds, and mixtures with effective atomic numbers, $Z_e$, in the range of $5-30$ since they are among the most
commonly occurring materials that are sensitive to X-rays in 
our chosen energy range.
Our choice of object's thicknesses was such that
the maximum attenuation value for the low-energy
DECT spectrum was in the range of $0.5-5.5$.
Outside this attenuation range, 
it is typically difficult to extract information from X-ray measurements due to the presence of noise. 
Such a wide selection of materials and 
attenuations is useful to demonstrate
the trends in performance as a function 
of $Z_e$ and object thicknesses.

We simulated fan-beam X-ray DECT data
at $256$ different view angles and $256$ pixel bins. 
The low-energy X-ray spectrum is generated
using a $100$kV bremsstrahlung X-ray source 
and a $2$mm Aluminum filter.
The high-energy X-ray spectrum is generated
using a $160$kV bremsstrahlung X-ray source,
a $2$mm aluminum filter, and a $2$mm copper filter.
DECT data is simulated for a circular object
that occupies a $224\times 224$ cross-sectional
pixel area within the full field-of-view of $256\times 256$. 
The source-to-object (SOD) distance
and source-to-detector (SDD) distance were 
chosen to be $500$mm and $1000$mm respectively.   
The diameter of the object was adjusted such that
we obtained maximum attenuation values of $0.5$ to $5.5$
in increments of $1.0$ for the low-energy spectrum.
For each attenuation value, given the atomic composition of the object's material, we computed the diameter that produces the desired low-energy attenuation along the X-ray path passing through the center of the object.
Then, the pixel size for the object was obtained by dividing this diameter by $224$ (number of pixels occupied by the object along any direction).
We assume a Perkin-Elmer (PE) detector for the
detector spectral response.  
For generating simulated data, 
we used analytic ray-tracing through geometric solids that is implemented in LTT \cite{champley_livermore_2022}. 
 We add $0.1\%$ noise to the
 measurements in transmission space.

For quantitative comparison of $Z_e$ and 
$\rho_e$ reconstructions, 
we utilize two forms of error metrics.
The first error metric estimates the bias in the reconstructed values by computing the relative error for either $Z_e$ or $\rho_e$ as,
\begin{align}
RE_{SIRZ-K}^x & =  100\times\frac{1}{x^{GT}}\lt[\lt(\frac{1}{\lt|\mathcal{M}\rt|}\sum_{j\in \mathcal{M}} x_{j}\rt) - x^{GT}\rt], \label{eq:relerr}
\end{align}
where $K\in \lt(2, 3\rt)$ is used to denote either SIRZ-2 or SIRZ-3 methods, 
$x\in\lt(\rho_e, Z_e\rt)$ denotes either $\rho_e$ or $Z_e$,
$\mathcal{M}$ is the set of all pixel indices
in the interior of the object that excludes the object boundaries, $\lt|\mathcal{M}\rt|$ is the cardinality (number of elements) of the set $\mathcal{M}$, and $x^{GT}$ is the ground-truth value for $x_j$.
The second error metric is the relative root mean squared error (RMSE), which is defined as, 
\begin{align}
RMSE_{SIRZ-K}^{x} & =  100\times\frac{1}{x^{GT}}\sqrt{\frac{1}{\lt|\mathcal{M}\rt|}\sum_{j\in \mathcal{M}} \lt(x_j - x^{GT}\rt)^2}, \label{eq:relrmse}
\end{align}
The relative RMSE metric combines both the bias 
and standard deviation of the difference between the reconstruction and the ground-truth. 

In Fig. \ref{fig:simres}, we compare the performance of SIRZ-2 and SIRZ-3
for a disc of copper (Cu) with a diameter of $4mm$ and a maximum attenuation of $4.5$ at the low-energy spectrum. 
Fig. \ref{fig:simres} (a, b) show the ground-truth
$\lt(Z_e, \rho_e\rt)$ images 
and Fig. \ref{fig:simres} (c, d) show the low-energy
and high-energy sinograms.
SIRZ-2 and SIRZ-3 reconstructions are shown 
in  Fig. \ref{fig:simres} (e, f) and Fig. \ref{fig:simres} (g, h) respectively.
We can see that both the relative error and the relative RMSE
reduces with SIRZ-3 when compared to SIRZ-2.
By comparing Fig. \ref{fig:simres} (e,f)
with Fig. \ref{fig:simres} (a,b), we
see that SIRZ-2 over estimates the 
$Z_e$ and under estimates the $\rho_e$ reconstructions.
Alternatively, by comparing Fig. \ref{fig:simres} (g,h) with Fig. \ref{fig:simres} (a,b), we see that SIRZ-3 reduces the bias in $Z_e$ and $\rho_e$ estimates.

In Fig. \ref{fig:Zeerrvsmatatten} and \ref{fig:rhoeerrvsmatatten},
we compare the performance of SIRZ-2 and SIRZ-3
using the relative error metric (equation \eqref{eq:relerr}).
We compare the relative errors in $Z_e$ and $\rho_e$ for various materials
with $Z_e$ in the range of $5-30$ and 
maximum low-energy attenuation values, denoted as $y_L^{max}$, in the range of $0.5-5.5$.
Note that $y_L^{max}$ is the maximum value
of low-energy attenuation (modeled using \eqref{eq:forwmodL}) in the absence of noise.
The rows of the heat map are sorted
according to ascending values of $Z_e$.
In Fig. \ref{fig:Zeerrvsmatatten} and \ref{fig:rhoeerrvsmatatten}, pure element materials are specified by their chemical formula,
PVC ($Z_e^{GT} = 14.07$)\footnote{$Z^{GT}_e$ denotes the ground-truth value for $Z_e$.} denotes polyvinyl chloride,
RbBr is $19\%$ RbBr salt solution in water ($Z_e^{GT}=20.05, \rho^{GT}_e=0.614\times10^{-3}$electrons$\times$mol/mm$^3$), 
Ti64 ($Z_e^{GT}=21.67$) is an alloy of Al ($2.8\%$), Ti ($23.9\%$), and V, 
Ti5553 ($Z_e^{GT}=23.73$) is an alloy of Al ($3.6\%$), Ti ($32.8\%$), V ($1.9\%$), Cr ($1.1\%$), and Mo, 
and brass is an alloy of Cu ($1.5\%$) and Zn. 

The reduction in $Z_e$ relative error obtained using SIRZ-3 
is quantified in Fig. \ref{fig:Zeerrvsmatatten} (c),
which shows the difference between the absolute values of the
relative errors using SIRZ-2 (Fig. \ref{fig:Zeerrvsmatatten} (a)) and SIRZ-3 (Fig. \ref{fig:Zeerrvsmatatten} (b)). 
A positive value in Fig. \ref{fig:Zeerrvsmatatten} (c)
indicates that the SIRZ-2 relative error is larger 
in magnitude than the SIRZ-3 relative error.
For LDPE, we observe that the SIRZ-3 absolute\footnote{Henceforth, relative error refers to the absolute value of the relative error.} relative error is more than $1.5\%$ lower than the SIRZ-2 relative error when $y_L^{max}\leq 2.5$. 
Beyond LDPE, at low values for $Z_e$ of up to $Z_e=20$ (calcium), 
we see that both SIRZ-2 and SIRZ-3 errors are within $1\%$. 
Hence, there does not seem to be a benefit to using SIRZ-3 over SIRZ-2
since all errors are below $1\%$.
With RbBr salt solution, we also do not see any apparent advantage of SIRZ-3 over SIRZ-2.
For titanium ($Z_e=22$) and its alloys,
we observe a marginal improvement in $Z_e$ performance with SIRZ-3 ($\approx 0\%-2\%$ reduction in relative error).
For materials such as iron (Fe), copper (Cu), brass, and zinc (Zn) with $Z_e\geq 26$, 
the SIRZ-2 relative errors rise beyond $14\%$ when $y_L^{max}\geq 3.5$.
However, we observe that the SIRZ-3
relative error is always less than $2.5\%$ for these materials.
From Fig. \ref{fig:Zeerrvsmatatten}, 
we observe a trend where
the performance of SIRZ-3 over SIRZ-2
improves with increasing values for both $y_L^{max}$
and $Z_e$ when $Z_e$ is greater than $\approx 22$. 
The $Z_e$ relative errors with SIRZ-3 are almost always within $2.5\%$ (except RbBr) irrespective of the $Z_e$ value.
However, the SIRZ-2 relative error increases
to more than $25\%$ as the value for $Z_e$ 
and $y_L^{max}$ is increased. 
In Fig. \ref{fig:Zeerrvsmatatten}, 
SIRZ-3 improves upon SIRZ-2 in $80\%$ of the simulated cases where the absolute value of the $Z_e$ relative errors with either SIRZ-2 or SIRZ-3 is greater than $1\%$, . 

 The reduction in $\rho_e$ relative error obtained using SIRZ-3 
is quantified in Fig. \ref{fig:rhoeerrvsmatatten} (c),
which shows the difference between the absolute
relative errors using SIRZ-2 (Fig. \ref{fig:rhoeerrvsmatatten} (a)) and SIRZ-3 (Fig. \ref{fig:rhoeerrvsmatatten} (b)). 
A positive value in Fig. \ref{fig:rhoeerrvsmatatten} (c)
indicates that the SIRZ-2 relative error is larger 
in magnitude than the SIRZ-3 relative error.
At low values for $Z_e$ of up to PVC, 
there does not seem to be a benefit to using SIRZ-3 since the relative errors in $\rho_e$ are always under $1\%$.
For calcium, SIRZ-3 reduces
the $\rho_e$ relative error
by more than $1\%$ when $y^{max}_L\geq 2.5$. 
SIRZ-3 does not seem to offer any improvements for RbBr salt solution. 
For Ti and its alloys, the reduction 
in SIRZ-3 relative error when compared to SIRZ-2  is in the range of $0.5-4\%$ when $y_L^{max}\geq 2.5$.
For materials such as Fe, Cu, Brass, and Zn with
$Z_e\geq 26$, we observe more than $20\%$
reduction in relative error using SIRZ-3 when $y_L^{max}\geq 2.5$. 
From Fig. \ref{fig:rhoeerrvsmatatten}, 
we observe a general trend where
the $\rho_e$ performance of SIRZ-3 over SIRZ-2
improves with increasing values for $y_L^{max}$
and $Z_e$ when $Z_e$ is greater than $\approx 22$. 
While the SIRZ-2 relative errors in $\rho_e$ reach very large 
values of more than $40\%$, the SIRZ-3
relative errors in $\rho_e$ are almost always under $5\%$ (except RbBr).
In Fig. \ref{fig:rhoeerrvsmatatten}, 
SIRZ-3 improves upon SIRZ-2 in $87\%$ of the simulated cases where the absolute value of the $\rho_e$ relative errors with either SIRZ-2 or SIRZ-3 is greater than $1\%$.

\section{Conclusions}
We presented the SIRZ-3 method for reconstruction 
of the effective atomic number ($Z_e$) and electron density ($\rho_e$) of an object imaged using
X-ray dual-energy computed tomography (DECT).
We presented a differentiable forward model that
expresses the DECT data as a direct analytical 
function of the unknown $\lt(Z_e, \rho_e\rt)$.
SIRZ-3 uses numerical optimization to solve for 
$\lt(Z_e, \rho_e\rt)$ such that the output of the forward
model is close to the measured data. 
In general, compared to SIRZ-2, the magnitude of performance improvement with SIRZ-3 increases 
as a function of increasing $Z_e$. 
For materials such as iron, copper, brass, and zinc with $Z_e\geq 26$, we show that SIRZ-3 provides more than $10\%$ reduction in $Z_e$ relative error
and more than $20\%$ reduction in $\rho_e$ relative error when the maximum low-energy attenuation is greater than $2.5$. 
The large errors for SIRZ-2 at high $Z_e$
is due to approximations made in the dual-energy decomposition and $\rho_e/Z_e$ conversion steps.
Due to the absence of these 
approximations, SIRZ-3 performs better than SIRZ-2.
We tested materials with $Z_e$ values in the range
of $5-30$ and low-energy attenuation values
in the range of $0.5-5.5$.
During $Z_e$ reconstruction, SIRZ-3 improves upon SIRZ-2 in $80\%$ of our simulated cases where the absolute value of the $Z_e$ relative errors with either SIRZ-2 or SIRZ-3 is greater than $1\%$.
During $\rho_e$ reconstruction, SIRZ-3 improves upon SIRZ-2 in $87\%$ of our simulated cases where the absolute value of the $\rho_e$ relative errors with either SIRZ-2 or SIRZ-3 is greater than $1\%$.

\section{Acknowledgments}
LLNL-CONF-832227. This work was performed under the auspices of the U.S. Department of Energy by Lawrence Livermore National Laboratory under Contract DE-AC52-07NA27344.
LDRD 21-FS-013 was used for this project.

\bibliography{main} 
\bibliographystyle{spiebib} 

\end{document}